\documentclass[reprint,amsmath,amssymb]{revtex4-2}
\usepackage[utf8]{inputenc}
\usepackage[T1]{fontenc}
\usepackage{graphicx}
\usepackage[svgnames]{xcolor}
\usepackage{dcolumn}
\usepackage{bm}
\usepackage{enumerate}
\usepackage{hyperref}
\usepackage{wasysym}
\hypersetup{
 colorlinks=true,%
 linkcolor=DarkSlateBlue,
 citecolor=DarkSlateBlue,
 urlcolor=DarkSlateBlue,
}
\begin{document}

\preprint{}

\title{Emergence of the Cotton tensor for describing gravity}

\author{Junpei Harada}
 \email{jharada@hoku-iryo-u.ac.jp}
\affiliation{Health Sciences University of Hokkaido, Japan}

\date{May 31, 2021}

\begin{abstract}
It is shown that the Cotton tensor can describe the effects of gravity beyond general relativity. Any solution of the Einstein equations with or without the cosmological constant satisfies the field equations described by the Cotton tensor. It implies that the cosmological constant is an integration constant. A vacuum of a theory is  represented by the vanishing of the Cotton tensor, rather than the vanishing of the Ricci tensor. An exact Schwarzschild-like solution for a static and spherically symmetric source is discovered. Although the field equations involve the third order of derivative, it is found that they reduce to the second-order differential equations due to a variational principle.
\end{abstract}
\maketitle

\section*{Introduction}
The Cotton curvature tensor, named after \'{E}mile Cotton~\cite{Cotton:1899}, is a rank-3 tensor that describes the curvature of a Riemannian manifold of dimension $n$. For $n<3$, it identically vanishes. For $n=3$, it vanishes if and only if a manifold is conformally flat. The Cotton tensor has mainly been used to investigate the three-dimensional manifolds~\cite{Cotton:1899} and three-dimensional gravity~\cite{Jackiw:2004qm,Deser:2004wd}. However, as we will see below, it is found that the Cotton tensor can describe the effects of gravity beyond general relativity in four-dimensional spacetime.

In this Letter it is shown that the Cotton tensor emerges as describing the gravitational field equations. This description has a number of advantages. First, any solution of the Einstein equations {\it with or without } the cosmological constant satisfies the given field equations. It implies that the cosmological constant is an integration constant---there is no need to add it to the field equations or to the action. Second, a vacuum in such a theory is represented by the vanishing of the Cotton tensor, rather than the vanishing of the Ricci tensor. This implies that the field equations in vacuum have more solutions than general relativity---it will be shown that any metric of the Ricci-flat manifolds, the Einstein manifolds, or the conformally flat manifolds is a trivial vacuum solution. As a nontrivial solution, an exact Schwarzschild-like metric for a static and spherically symmetric source is discovered. It is the first discovered nontrivial exact solution of the given field equations. Third, it is found that the field equations described by the Cotton tensor can be derived from the Weyl action of conformal gravity, by varying the action with respect to the connection keeping the metric fixed. Although the field equations involve the third order of derivative, they reduce to the second-order differential equations due to such a variational principle. 

Throughout this Letter we work in $n=4$ dimensions. The usual Levi-Civita connection is adopted to defined the covariant derivative, though a choice is arbitrary. The signature of metric is $(-,+,+,+)$. 

\section*{Gravitational field equations}
In electromagnetism, the Maxwell's equations in vacuum are written as $\partial_\mu F^{\mu\nu}=0$, where $\partial_\mu F^{\mu\nu}$ denotes the divergence of a linear curvature. In the case of gravity, there are four possible terms as
\begin{subequations}
\begin{eqnarray}
	&&\nabla_\mu R^{\mu\nu\rho\sigma},\label{eq:der_curv1}\\
	&&(g^{\mu\rho}g^{\nu\sigma} - g^{\nu\rho}g^{\mu\sigma})\nabla_\mu R,\label{eq:der_curv2}\\
	&&\nabla_\mu (g^{\mu\rho}R^{\nu\sigma} - g^{\nu\rho}R^{\mu\sigma} - g^{\mu\sigma}R^{\nu\rho} + g^{\nu\sigma}R^{\mu\rho}),\label{eq:der_curv3}\\	
	&&\nabla_\mu C^{\mu\nu\rho\sigma},\label{eq:der_curv4}
\end{eqnarray}
\end{subequations}
where $\nabla_\mu$ denotes the covariant derivative, and $C_{\mu\nu\rho\sigma}$ is the Weyl curvature tensor~\cite{Weyl:1918pdp} which is defined by 
\begin{eqnarray}
	C_{\mu\nu\rho\sigma} 
	:=&& R_{\mu\nu\rho\sigma} +\frac{1}{6}(g_{\mu\rho}g_{\nu\sigma} - g_{\nu\rho}g_{\mu\sigma})R\nonumber\\
	-&& \frac{1}{2}(g_{\mu\rho}R_{\nu\sigma} - g_{\nu\rho}R_{\mu\sigma} - g_{\mu\sigma}R_{\nu\rho} + g_{\nu\sigma}R_{\mu\rho}).
	\label{eq:Weyltensor}
\end{eqnarray}
It should be also noted that the following Bianchi identity is satisfied (this form may be unfamiliar, but it is useful),
\begin{eqnarray}
	&&\nabla_\mu R^{\mu\nu\rho\sigma}
	 +\frac{1}{2}(g^{\mu\rho}g^{\nu\sigma} - g^{\nu\rho}g^{\mu\sigma})\nabla_\mu R \nonumber \\
	&& -\nabla_\mu\left(g^{\mu\rho}R^{\nu\sigma} - g^{\nu\rho}R^{\mu\sigma} - g^{\mu\sigma}R^{\nu\rho} + g^{\nu\sigma}R^{\mu\rho}\right)=0.\label{eq:Bianchi}
\end{eqnarray}

Four possible terms~\eqref{eq:der_curv1}--\eqref{eq:der_curv4} are not linearly independent due to Eqs.~\eqref{eq:Weyltensor} and~\eqref{eq:Bianchi}, and therefore the gravitational analog of $\partial_\mu F^{\mu\nu}$ should be a linear combination of arbitrary  two terms of Eqs.~\eqref{eq:der_curv1}--\eqref{eq:der_curv4}. If we choose  \eqref{eq:der_curv1} and~\eqref{eq:der_curv2} as two independent terms, then the gravitational analog of $\partial_\mu F^{\mu\nu}$ can be written in the form, 
\begin{equation}
	a \nabla_\mu R^{\mu\nu\rho\sigma} + b (g^{\mu\rho}g^{\nu\sigma} - g^{\nu\rho}g^{\mu\sigma})\nabla_\mu R.
\end{equation}
The coefficients $a$ and $b$ will be determined soon. We regard this as the left-hand side of the gravitational field equations. The right-hand side of field equations can be similarly written. There are {\it only} two possible terms as
\begin{subequations}
\begin{eqnarray}
	&&\nabla_\mu (g^{\mu\rho}T^{\nu\sigma} - g^{\nu\rho}T^{\mu\sigma} - g^{\mu\sigma}T^{\nu\rho} + g^{\nu\sigma}T^{\mu\rho}),\label{eq:der_em1}\\
	&&(g^{\mu\rho}g^{\nu\sigma} - g^{\nu\rho}g^{\mu\sigma})\nabla_\mu T,\label{eq:der_em2}
\end{eqnarray}
\end{subequations}
where $T_{\mu\nu}$ is the energy-momentum tensor and $T=T^\mu_\mu$. Therefore, the field equations can be written as
\begin{eqnarray}
	\lefteqn{
	a \nabla_\mu R^{\mu\nu\rho\sigma} + b (g^{\mu\rho}g^{\nu\sigma} - g^{\nu\rho}g^{\mu\sigma})\nabla_\mu R}\nonumber\\
	&&=c \nabla_\mu (g^{\mu\rho}T^{\nu\sigma} - g^{\nu\rho}T^{\mu\sigma} - g^{\mu\sigma}T^{\nu\rho} + g^{\nu\sigma}T^{\mu\rho})\nonumber\\
	&&+ d (g^{\mu\rho}g^{\nu\sigma} - g^{\nu\rho}g^{\mu\sigma})\nabla_\mu T,
	\label{eq:GFE1}
\end{eqnarray}
where $c$ and $d$ are coefficients. 

Let us now determine the coefficients $a$, $b$, $c$, and $d$. Multiplying Eq.~\eqref{eq:GFE1} by $g_{\nu\sigma}$, we find 
\begin{equation}
	(a+6b)\nabla_\mu R^{\mu\rho} = 2c\nabla_\mu T^{\mu\rho} + (c+3d)\nabla_\mu (g^{\mu\rho}T),
	\label{eq:GFE1_g}
\end{equation}
where we have used the identity $g^{\mu\nu}\nabla_\mu R=2\nabla_\mu R^{\mu\nu}$. 
We now require that the conservation law of the energy momentum ($\nabla_\mu T^{\mu\rho}=0$) is identically satisfied, and then from Eq.~\eqref{eq:GFE1_g}, we obtain 
\begin{subequations}
\begin{eqnarray}
	&&a + 6b = 0,\label{eq:coef1}\\
	&&c+3d=0.\label{eq:coef2}
\end{eqnarray}
\end{subequations}
We also require that any solution of the Einstein equations satisfies Eq.~\eqref{eq:GFE1}. Inserting $T_{\mu\nu}=R_{\mu\nu} - Rg_{\mu\nu}/2$ (where we set $8\pi G=1$, $G$ is the Newton's constant) and $T=-R$ into the right-hand side of Eq.~\eqref{eq:GFE1}, we find that the right-hand side of Eq.~\eqref{eq:GFE1} is written as
\begin{equation}
	c \nabla_\mu R^{\mu\nu\rho\sigma} - \left(\frac{c}{2}+d\right) (g^{\mu\rho}g^{\nu\sigma} - g^{\nu\rho}g^{\mu\sigma})\nabla_\mu R,
	\label{eq:rhs}
\end{equation}
where the Bianchi identity~\eqref{eq:Bianchi} has been used. 
Comparing Eq.~\eqref{eq:rhs} with the left-hand side of Eq.~\eqref{eq:GFE1}, we obtain
\begin{subequations}
\begin{eqnarray}
	&&a = c,\label{eq:coef3}\\
	&&b=-\left(\frac{c}{2}+d\right).\label{eq:coef4}
\end{eqnarray}
\end{subequations}
From Eqs.~\eqref{eq:coef1}, \eqref{eq:coef2}, \eqref{eq:coef3}, and \eqref{eq:coef4} (three of them are linearly independent), the coefficients $a$, $b$, $c$, and $d$ are determined as 
\begin{equation}
	a=1,\quad
	b=-\frac{1}{6},\quad
	c=1,\quad
	d=-\frac{1}{3},\quad
	\label{eq:coef}
\end{equation}
where we set $a=1$ as a normalization. Thus, we obtained Eq.~\eqref{eq:GFE1} with Eq.~\eqref{eq:coef} as the field equations, but it is now convenient to rewrite them in a simple form as follows.

It is convenient to define the rank-4 tensor by
\begin{equation}
	 G^{\mu\nu\rho\sigma} := R^{\mu\nu\rho\sigma} -\frac{1}{6} (g^{\mu\rho}g^{\nu\sigma} - g^{\nu\rho}g^{\mu\sigma})R.
	 \label{eq:G4tensor}
\end{equation}
Multiplying $G^{\mu\nu\rho\sigma}$ by $g_{\nu\sigma}$, we find that
\begin{equation}
	g_{\nu\sigma}G^{\mu\nu\rho\sigma} = R^{\mu\rho} - \frac{1}{2}g^{\mu\rho}R =: G^{\mu\rho},
\end{equation}
where $G^{\mu\rho}$ is the Einstein tensor. It is also convenient to define the rank-4 tensor by
\begin{eqnarray}
	T^{\mu\nu\rho\sigma}:=&&
	\frac{1}{2}(g^{\mu\rho}T^{\nu\sigma} - g^{\nu\rho}T^{\mu\sigma} - g^{\mu\sigma}T^{\nu\rho} + g^{\nu\sigma}T^{\mu\rho})\nonumber\\
	&&-\frac{1}{6} (g^{\mu\rho}g^{\nu\sigma} - g^{\nu\rho}g^{\mu\sigma})T,
	\label{eq:T4tensor}
\end{eqnarray}
where an overall normalization of Eq.~\eqref{eq:T4tensor} has been chosen so that the following relation is satisfied,
\begin{equation}
	g_{\nu\sigma}T^{\mu\nu\rho\sigma} = T^{\mu\rho}.
\end{equation}
Therefore, Eq.~\eqref{eq:GFE1} with Eq.~\eqref{eq:coef} can be written as
\begin{equation}
	\nabla_\mu G^{\mu}{}_{\nu\rho\sigma} = 16\pi G \nabla_\mu T^{\mu}{}_{\nu\rho\sigma},\label{eq:GFE4}
\end{equation}
where $8\pi G$ is explicitly written, and we have
\begin{equation}
	\nabla_\mu T^{\mu}{}_{\nu\rho\sigma}
	=\frac{1}{2} (\nabla_\rho T_{\nu\sigma} - \nabla_\sigma T_{\nu\rho} )
	-\frac{1}{6} (g_{\nu\sigma}\nabla_\rho T - g_{\nu\rho}\nabla_\sigma T).
\end{equation}

It is even more convenient to rewrite Eq.~\eqref{eq:GFE4} as follows. Using Eq.~\eqref{eq:G4tensor}, we find 
\begin{eqnarray}
	\nabla_\mu G^{\mu}{}_{\nu\rho\sigma} 
	&&=\nabla_\mu R^{\mu}{}_{\nu\rho\sigma} -\frac{1}{6}(g_{\nu\sigma}\nabla_\rho R - g_{\nu\rho} \nabla_\sigma R ),\nonumber\\
	&&=\nabla_\rho R_{\nu\sigma} - \nabla_\sigma R_{\nu\rho}-\frac{1}{6}(g_{\nu\sigma}\nabla_\rho R - g_{\nu\rho}\nabla_\sigma R ),	\nonumber\\
	&&=:C_{\nu\rho\sigma},\label{eq:VGFE2}
\end{eqnarray}
where the identity $\nabla_\mu R^{\mu}{}_{\nu\rho\sigma}  = \nabla_\rho R_{\nu\sigma} - \nabla_\sigma R_{\nu\rho}$ has been used, and $C_{\nu\rho\sigma}$ is called the Cotton tensor that satisfies 
\begin{subequations}
\begin{eqnarray}
	&&\nabla^\nu C_{\nu\rho\sigma} = 0,\\
	&&g^{\nu\rho}C_{\nu\rho\sigma} = 0,\\	
	&&C_{\nu\rho\sigma} + C_{\nu\sigma\rho} = 0,\\
	&&C_{\nu\rho\sigma} + C_{\rho\sigma\nu} + C_{\sigma\nu\rho} = 0.
\end{eqnarray}
\end{subequations}
It should be noted that $C_{\mu\nu\rho\sigma}$ denotes the Weyl tensor, and $C_{\nu\rho\sigma}$ denotes the Cotton tensor. 

\begin{figure}[b]
	\includegraphics[width=85mm]{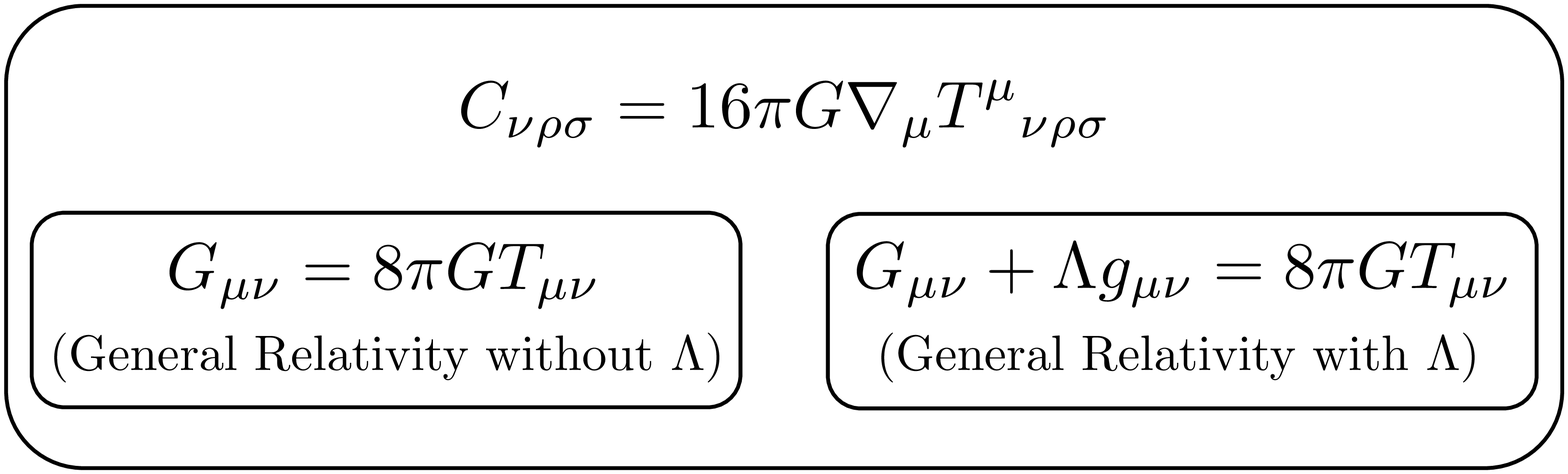}
	\caption{\label{fig:FE} The solutions of Einstein equations with or without the cosmological constant are subsets of those of Eq.~\eqref{eq:GFE5}. Whenever the Einstein equations (with or without the cosmological constant) are satisfied, Eq.~\eqref{eq:GFE5} is also satisfied. However, its inverse is not true---even if Eq.~\eqref{eq:GFE5} is satisfied, the Einstein equations are {\it not} necessarily satisfied.}
\end{figure}

Consequently, we obtain the field equations of gravity,
\begin{equation}
	C_{\nu\rho\sigma}= 16\pi G \nabla_\mu T^{\mu}{}_{\nu\rho\sigma}.\label{eq:GFE5}
\end{equation}
Multiplying Eq.~\eqref{eq:GFE5} by $g^{\nu\sigma}$, we can quickly confirm the conservation law of the energy-momentum tensor as 
\begin{equation}
	g^{\nu\sigma} C_{\nu\rho\sigma} = \nabla_\mu G^\mu{}_{\rho} =  16\pi G \nabla_\mu T^\mu{}_{\rho}=0.
\end{equation}
Here we should present some remarks. First, any solution of the Einstein equations ($G_{\mu\nu}=8\pi G T_{\mu\nu}$) satisfies Eq.~\eqref{eq:GFE5}. Furthermore, it is obvious that any solution of the Einstein equations with the cosmological constant ($G_{\mu\nu} + \Lambda g_{\mu\nu}=8\pi G T_{\mu\nu}$) satisfies Eq.~\eqref{eq:GFE5}. This implies that the cosmological constant is an integration constant. Of course, Eq.~\eqref{eq:GFE5} has other solutions which are not solutions of the Einstein equations---Eq.~\eqref{eq:GFE5} has more information than general relativity. It should be noted that the Einstein equations need {\it not} be satisfied in Eq.~\eqref{eq:GFE5}. These are shown in Fig.~\ref{fig:FE}.

\section*{Exact vacuum solutions}
In a vacuum, Eq.~\eqref{eq:GFE5} yields
\begin{equation}
	C_{\nu\rho\sigma}
	=0. \label{eq:VGFE1}
\end{equation}
This should be regarded as a generalization of $R_{\mu\nu}=0$ of general relativity. 
It is clear from Eq.~\eqref{eq:VGFE2} that if $R_{\mu\nu}$ is zero, then $C_{\nu\rho\sigma}$ vanishes too, so any vacuum solution of the Einstein equations is a solution of Eq.~\eqref{eq:VGFE1}. We also find from Eq.~\eqref{eq:VGFE2} that if $R_{\mu\nu}=\Lambda g_{\mu\nu}$, then $C_{\nu\rho\sigma}$ automatically vanishes. Thus, the de Sitter (or the anti--de Sitter) metric is a solution of Eq.~\eqref{eq:VGFE1}, though there is no cosmological constant in Eq.~\eqref{eq:VGFE1}. 

Furthermore, any metric of the locally conformally flat manifolds satisfies Eq.~\eqref{eq:VGFE1}. This can be understood as follows. Using the Bianchi identity~\eqref{eq:Bianchi}, we find 
\begin{equation}
	\nabla_\mu G^{\mu}{}_{\nu\rho\sigma}
	=2\nabla_\mu C^{\mu}{}_{\nu\rho\sigma}
	=C_{\nu\rho\sigma}.
	\label{eq:Weyl_derivative}
\end{equation}
This indicates that the divergence of $G^{\mu\nu\rho\sigma}$ and that of $C^{\mu\nu\rho\sigma}$ are equivalent (up to a factor two), though 
$G^{\mu\nu\rho\sigma}$ and $C^{\mu\nu\rho\sigma}$ are different. It is clear from Eq.~\eqref{eq:Weyl_derivative} that if $C_{\mu\nu\rho\sigma}$ is zero, then $C_{\nu\rho\sigma}$ vanishes too, so any metric of the locally conformally flat manifold $(C_{\mu\nu\rho\sigma}=0)$ satisfies Eq.~\eqref{eq:VGFE1}. 

Thus, it is concluded that any metric that satisfies
\begin{subequations}
\begin{eqnarray}
	R_{\mu\nu}&=&0,\qquad
	\mbox{(Ricci-flat manifold)}	\label{eq:Ricci-flat}\\
	R_{\mu\nu}&=&\Lambda g_{\mu\nu},\
	\mbox{(Einstein manifold)}	\label{eq:Einstein-manifold}\\
	C_{\mu\nu\rho\sigma}&=&0,\qquad
	\mbox{(conformally flat manifold)}	\label{eq:conformallyflat}
\end{eqnarray}
\end{subequations}
is a solution of Eq.~\eqref{eq:VGFE1}.

The vacuum equations ($C_{\nu\rho\sigma}=0$) have other solutions, which does not satisfy Eqs.~\eqref{eq:Ricci-flat}--\eqref{eq:conformallyflat}. Here we present such a nontrivial solution. We present a static, spherically symmetric exact solution, which may be regarded as a generalization of the Schwarzschild metric.

Let us focus on the Schwarzschild-like metric as
\begin{equation}
	ds^2 = - e^{\nu (r)} dt^2 + e^{-\nu (r)}dr^2+ r^2 d\Omega^2,
\end{equation}
where $d\Omega^2=d\theta^2 + \sin^2\theta d\phi^2$. Inserting this into Eq.~\eqref{eq:VGFE1}, we find that the component of $C_{\nu\rho\sigma}$ is given by
\begin{eqnarray}
	&&3e^{-\nu}C_{010}=- \frac{2}{r^3}\nonumber\\
	&&+\left(\nu^{\prime\prime\prime}+3\nu^\prime \nu^{\prime\prime}+{\nu^\prime}^3 + \frac{\nu^{\prime\prime}}{r}+\frac{{\nu^\prime}^2}{r}-\frac{2\nu^\prime}{r^2}+\frac{2}{r^3}\right)e^\nu,\nonumber\\
	\label{eq:G010}
\end{eqnarray}
where a prime denotes the derivative with respect to $r$. Other components of $C_{\nu\rho\sigma}$ vanish except for $C_{212}$ and $C_{313}$, which are proportional to $C_{010}$. The substitution
\begin{equation}
	y(r)=\left(\nu^\prime + \frac{1}{r} \right)e^\nu
	\label{eq:substitution}
\end{equation}	
enables us to rewrite Eq.~\eqref{eq:G010} in a simple form as
\begin{equation}
	3e^{-\nu}C_{010} = y^{\prime\prime} - \frac{2}{r^3}. 
	\label{eq:2nd_derivative}
\end{equation}
Then, the vacuum equation $C_{010}=0$ is solved as
\begin{equation}
	y = \frac{1}{r} + c_1 r + c_2,
\end{equation}
where $c_1$ and $c_2$ are integration constants, and Eq.~\eqref{eq:substitution} is written as
\begin{equation}
	\left(\nu^\prime + \frac{1}{r}\right)e^\nu = \frac{1}{r} + c_1 r + c_2.
\end{equation}
This equation is easily solved as
\begin{equation}
	-g_{00} = 1/g_{11}=e^{\nu} = 1 + \frac{c_3}{r} + \frac{c_1}{3}r^2 + \frac{c_2}{2}r,
\end{equation}
where $c_3$ is an integration constant. 

If we rename three integration constants as $c_3 = -2M$, $c_1 = - \Lambda$, and $c_2 = 2\gamma$, then the solution is written as
\begin{equation}
	-g_{00} = 1/g_{11} = 1 - \frac{2M}{r} - \frac{\Lambda}{3}r^2 + \gamma r. 
	\label{eq:Sch_like}
\end{equation}
This is exact. This is nontrivial in the sense that it does not satisfy Eqs.~\eqref{eq:Ricci-flat}--\eqref{eq:conformallyflat}. This is the first discovered nontrivial exact solution of Eq.~\eqref{eq:VGFE1}. Furthermore, the linear and quadratic curvature invariants are given by
\begin{subequations}
\begin{eqnarray}
	R^{\mu\nu\rho\sigma}R_{\mu\nu\rho\sigma} 
	&=& \frac{48M^2}{r^6} +\frac{8\gamma^2}{r^2} - \frac{8\gamma \Lambda}{r} + \frac{8\Lambda^2}{3},
	\\
	R^{\mu\nu}R_{\mu\nu} 
	&=& \frac{10\gamma^2}{r^2} - \frac{12\gamma \Lambda}{r} + 4 \Lambda^2,\\
	R&=&-\frac{6\gamma}{r} + 4\Lambda,\\
	C^{\mu\nu\rho\sigma}C_{\mu\nu\rho\sigma}
	&=&\frac{48M^2}{r^6}.
\end{eqnarray}
\end{subequations}
We can confirm that these invariants reduces to those of the Schwarzschild--de Sitter metric in the $\gamma \rightarrow 0$ limit.

In Eq.~\eqref{eq:Sch_like}, the $\gamma r$ term cannot be obtained from the Einstein equations. When the $\gamma r$ term is negligible, Eq.~\eqref{eq:Sch_like} reduces to the Schwarzschild--de Sitter metric, and therefore it is consistent with the observational tests of general relativity in that case.
Furthermore, Eq.~\eqref{eq:Sch_like} is {\it approximately} equivalent to an exact solution of conformal gravity~\cite{Mannheim:1988dj,Mannheim:1990wi}. In conformal gravity, the values of $\gamma$ have been estimated~\cite{Mannheim:2005bfa,Varieschi:2008fc,Varieschi:2008va}. It has been pointed out that the $\gamma r$  term can potentially explain the galactic rotation curve without the need of dark matter~\cite{Mannheim:1988dj,Mannheim:1992vj,Mannheim:2005bfa}, but there is still controversy in astrophysics~\cite{Hobson:2021vy}. 

Here we point out about the possible role of the $\gamma r$ term in the Solar System. In the Solar System, Eq.~\eqref{eq:Sch_like} can be written as
\begin{equation}
	-g_{00} = 1/g_{11} = 1 - \frac{2GM_{\astrosun}}{rc^2} + \gamma r,
	\label{eq:Solar}
\end{equation}
where the Newton's constant $G$ and the speed of light $c$ are explicitly written, and we have ignored the cosmological $-\Lambda r^2/3$ term, which is assumed to be small enough on the scale of the Solar System. We find that 
\begin{align}
	\frac{\partial}{\partial r}\left(-\frac{c^2 g_{00}}{2}\right) = \frac{GM_{\astrosun}}{r^2} + \frac{\gamma c^2}{2},
\end{align}
where $GM_{\astrosun}/r^2$ represents the usual inverse square law of Newton's gravity, and $\gamma c^2/2$ represents an additional acceleration which is independent of $r$. Such a constant acceleration has been observed by the Pioneer 10 and Pioneer 11 spacecrafts at distances of 20--70 AU, and the reported value is $a_{\rm Pioneer}=8.7\times 10^{-10} \ {\rm m \ s^{-2}}$~\cite{Turyshev:2010yf}. If this is due to $\gamma c^2/2$, then we have $\gamma = 1.9 \times 10^{-26}$ m$^{-1}$. Although it has been reported that the observed anomalous acceleration can be explained by thermal effects~\cite{Turyshev:2012mc}, our gravity theory also gives a possible explanation.

When $M=0$ in Eq.~\eqref{eq:Sch_like} (or when $r$ is sufficiently large so that $-2M/r$ term can be ignored), the metric is conformally flat ($C_{\mu\nu\rho\sigma}=0$). In such a case, any metric that is conformal to Eq.~\eqref{eq:Sch_like} is a solution of Eq.~\eqref{eq:VGFE1}. This can be understood as follows. We can show that the Cotton tensor transforms under the local conformal transformation of the metric $g_{\mu\nu}(x) \rightarrow \Omega^2 (x) g_{\mu\nu}(x)$ as
\begin{equation}
	C_{\nu\rho\sigma} \rightarrow C_{\nu\rho\sigma} + \Omega^{-1} (\partial_\mu \Omega) C^\mu{}_{\nu\rho\sigma}.
	\label{eq:Cotton_trans}
\end{equation}
Thus, the Cotton tensor is locally conformally invariant if and only if the Weyl tensor vanishes. Therefore, it is concluded that any metric which is conformal to Eq.~\eqref{eq:Sch_like} (with $M=0$) satisfies Eq.~\eqref{eq:VGFE1}. 

As explicitly shown in Eq.~\eqref{eq:G010}, the vacuum equations ($C_{\nu\rho\sigma}=0$) involve the third order of derivative. However, as shown in Eq.~\eqref{eq:2nd_derivative}, they reduce to the second-order differential equations. This is a generic property of our theory. As we will see below, this is originated from the variational principle that yields $C_{\nu\rho\sigma}=0$.

\section*{Action and variational principle}
The Weyl tensor in Eq.~\eqref{eq:Weyl_derivative} aims our attention to the following Weyl action,
\begin{eqnarray}
 I_{\rm W} &=& \frac{1}{2} \int d^4 x \sqrt{-g} C^{\mu\nu\rho\sigma}C_{\mu\nu\rho\sigma}\nonumber\\
 &=&\frac{1}{4}\int d^4 x \sqrt{-g}\left(R^{\mu\nu\rho\sigma}R_{\mu\nu\rho\sigma}-\frac{1}{3}R^2\right),
 \label{eq:Weylaction}
\end{eqnarray}
where $g=\det (g_{\mu\nu})$, and we have used the identity,
\begin{equation}
C^{\mu\nu\rho\sigma}C_{\mu\nu\rho\sigma}=R^{\mu\nu\rho\sigma}R_{\mu\nu\rho\sigma}-2R^{\mu\nu}R_{\mu\nu}+\frac{1}{3}R^2,
\end{equation}
and we have omitted the topological invariant from the action. The Weyl action~\eqref{eq:Weylaction} is a unique gravity action that is invariant under the local conformal transformation $g_{\mu\nu}(x) \rightarrow \Omega^2 (x) g_{\mu\nu}(x)$. We will show that the Weyl action~\eqref{eq:Weylaction} yields $C_{\nu\rho\sigma}=0$, but before showing it, we present some remarks on the conformal gravity.

The gravity theory with the Weyl action~\eqref{eq:Weylaction} is usually known as the conformal gravity~\cite{Mannheim:1988dj}. However, our gravity theory is different from conformal gravity, though the action coincides. The difference is originated from not the action but the variational principle; in conformal gravity, one varies the Weyl action~\eqref{eq:Weylaction} with respect to the metric, and it yields the rank-2 tensor equations, which are described by the Bach tensor~\cite{Bach_1922,Mannheim:1988dj,THooft:2015jcw}. Thus, a vacuum of conformal gravity is represented by the vanishing of the Bach tensor. In contrast, Eq.~\eqref{eq:VGFE1} is described by the Cotton tensor. This implies that the variational principle should be different from that of conformal gravity. 

The variational principle that yields $C_{\nu\rho\sigma}=0$ is as follows. Varying the Weyl action~\eqref{eq:Weylaction} with respect to the connection keeping the metric fixed, we find 
\begin{eqnarray}
	\delta I_{\rm W} &=& \frac{1}{2} \int d^4 x \sqrt{-g} ( R_\mu{}^{\nu\rho\sigma}\delta R^\mu{}_{\nu\rho\sigma} - \frac{1}{3}Rg^{\mu\nu}\delta R_{\mu\nu}) \nonumber\\
	&=& \int d^4 x \sqrt{-g} C^{\nu\rho\sigma}\delta \Gamma_{\sigma\nu\rho},
	\label{eq:variationWeylaction}
\end{eqnarray}
where $\delta \Gamma_{\sigma\nu\rho}:=g_{\sigma\lambda}\delta \Gamma^\lambda{}_{\nu\rho}$; we have used the identity
\begin{equation}
	\delta R^\mu{}_{\nu\rho\sigma}
	= \nabla_\rho (\delta \Gamma^\mu{}_{\sigma\nu}) - \nabla_{\sigma}(\delta \Gamma^\mu{}_{\rho\nu}),
	\label{eq:Palatini}
\end{equation}
and the boundary condition $\delta \Gamma^\rho{}_{\mu\nu}=0$ at the boundary is imposed. Thus, we find that $\delta I_{\rm W}=0$ yields $C^{\nu\rho\sigma}=0$. Thus, it is found that the local conformal symmetry of the action~\eqref{eq:Weylaction} can be broken in the field equations, as shown in Eq.~\eqref{eq:Cotton_trans}, due to a variational principle.

The variation of the connection (with keeping in mind that the metric fixed) is given by
\begin{equation}
	\delta \Gamma^\rho{}_{\mu\nu} 
	= \frac{1}{2}g^{\rho\lambda}
	\left(\delta(\partial_\mu g_{\nu\lambda}) + \delta (\partial_\nu g_{\mu\lambda}) - \delta (\partial_\lambda g_{\mu\nu})\right).
\end{equation}
Therefore, we are varying the action with respect to the derivative of the metric keeping the metric itself fixed. As a consequence of this variational principle, the field equations that involve the third order of derivative reduce to the second-order differential equations, as shown in Eq.~\eqref{eq:2nd_derivative}. This is a generic property of the theory, and it is a consequence of the variation presented here. 

Finally, we present the gravitational analog of the source term $J^\mu A_\mu$ in Maxwell's theory,
\begin{equation}
	I_{\rm source} = -8\pi G \int d^4 x \sqrt{-g}T^{\mu\nu\rho\sigma}R_{\mu\nu\rho\sigma},
	\label{eq:source}
\end{equation}
where $T^{\mu\nu\rho\sigma}$ is defined by Eq.~\eqref{eq:T4tensor}, and $T_{\mu\nu}$ is a source that does not include the connection. We find that the variation of the action with respect to the connection keeping the metric fixed yields, 
\begin{equation}
	\delta I_{\rm source}= -16 \pi G \int d^4 x \sqrt{-g} \nabla_\mu T^{\mu\nu\rho\sigma} \delta \Gamma_{\sigma\nu\rho}.
\end{equation}
Therefore, we find that $\delta I_{\rm W} + \delta I_{\rm source}=0$ yields Eq.~\eqref{eq:GFE5}. 

\section*{Conclusion}
It has been shown that the Cotton tensor can describe the effects of gravity beyond general relativity. A vacuum is represented by the vanishing of the Cotton tensor. An exact Schwarzschild-like solution has been discovered. This theory includes general relativity as part of itself, and has more information on gravity. It is a potentially viable theory of gravity, and it would be worth for further investigations.

\bibliography{references}

\end{document}